\documentclass[12pt,a4paper]{article}
\usepackage{a4wide}
\usepackage{amsthm,amssymb}
\usepackage{amsbsy,amsfonts,amsmath,amscd}
\usepackage{mathrsfs}

\usepackage{cite}
\usepackage{array}
\usepackage{epsfig,multicol}
\usepackage{verbatim}

\let\oldPhi=\Phi
\let\oldPsi=\Psi
\let\oldGamma=\Gamma
\let\oldDelta=\Delta
\let\oldSigma=\Sigma
\let\oldLambda=\Lambda
\let\oldTheta=\Theta
\let\oldPi=\Pi
\renewcommand{\Phi}{\mathnormal{\oldPhi}}
\renewcommand{\Psi}{\mathnormal{\oldPsi}}
\renewcommand{\Gamma}{\mathnormal{\oldGamma}}
\renewcommand{\Sigma}{\mathnormal{\oldSigma}}
\renewcommand{\Delta}{\mathnormal{\oldDelta}}
\renewcommand{\Theta}{\mathnormal{\oldTheta}}
\renewcommand{\Lambda}{\mathnormal{\oldLambda}}
\renewcommand{\Pi}{\mathnormal{\oldPi}}

\newcommand{\gen}[1]{\mathfrak{#1}}
\newcommand{\genY}[1]{\widehat{\mathfrak{#1}}}
\newcommand{\cwgen}[1]{\mathfrak{#1}}
\newcommand{\csgen}[1]{\mathfrak{#1}}

\newcommand{\idm}{\text{Id}}
\newcommand{\perm}{\mathcal{P}}

\newcommand{\qnum}[1]{\frac{q^{#1} - q^{-#1}}{q - q^{-1}}}
\newcommand{\qnumb}[1]{[#1]_q}

\newcommand{\cartnm}{A^{\alg{sl}(n|m)}}
\newcommand{\cartnn}{A^{\alg{gl}(n|n)}}
\newcommand{\carttnm}{\tilde{A}^{\alg{gl}(n|m)}}
\newcommand{\cartq}[1]{A_{#1}(q)}
\newcommand{\cartqb}{A(q)}

\newcommand{\csgh}[1]{\mathfrak{H}_{#1}}

\newcommand{\gfhp}[2]{\mathfrak{H}^+_{#1}(\lambda_{#2})}
\newcommand{\gfhm}[2]{\mathfrak{H}^-_{#1}(\lambda_{#2})}

\newcommand{\cwgp}[1]{\mathfrak{E}_{#1}}
\newcommand{\cwgm}[1]{\mathfrak{F}_{#1}}

\ifx\genfrac\sdflkaj

\else

\fi

\newcommand{\sprod}[2]{\left(#1,#2\right)}

\newcommand{\comm}[2]{[#1,#2]}
\newcommand{\adj}[3]{\left(\text{ad}_{#1}\right)^{#2}(#3)}

\newcommand{\scomm}[2]{[#1,#2\}}
\newcommand{\acomm}[2]{\{#1,#2\}}
\newcommand{\sacomm}[2]{\{#1,#2]}

\newcommand{\alg}[1]{\mathfrak{#1}}

\newcommand{\gltt}{\alg{gl}(2|2)}
\newcommand{\sln}{\alg{sl}(n)}
\newcommand{\slnm}{\alg{sl}(n|m)}
\newcommand{\glnm}{\alg{gl}(n|m)}
\newcommand{\slnn}{\alg{sl}(n|n)}
\newcommand{\glnn}{\alg{gl}(n|n)}

\newcommand{\sun}{\alg{su}(n)}
\newcommand{\Ya}[1]{\mathcal{Y}\left( #1\right )}
\newcommand{\DY}[1]{\mathcal{DY}\left( #1\right )}

\newcommand{\nln}{\nonumber\\}

\newcommand{\beq}{\begin{equation}}
\newcommand{\eeq}{\end{equation}}
\newcommand{\beqa}{\begin{eqnarray}}
\newcommand{\eeqa}{\end{eqnarray}}
\newcommand{\bal}{\begin{align}}
\newcommand{\eal}{\end{align}}

\def\[{\begin{equation}}
\def\]{\end{equation}}
\def\<{\begin{eqnarray}}
\def\>{\end{eqnarray}}

\makeatletter
\def\mr@ignsp#1 {\ifx\:#1\@empty\else #1\expandafter\mr@ignsp\fi}%
\newcommand{\multiref}[1]{\begingroup
\xdef\mr@no@sparg{\expandafter\mr@ignsp#1 \: }%
\def\mr@comma{}%
\@for\mr@refs:=\mr@no@sparg\do{\mr@comma\def\mr@comma{,}\ref{\mr@refs}}%
\endgroup}
\makeatother

\newcommand{\hypref}[2]{\ifx\href\asklfhas #2\else\href{#1}{#2}\fi}

\renewcommand{\eqref}[1]{(\multiref{#1})}

\ifx\href\asklfhas\newcommand{\href}[2]{#2}\fi

\title{The Yangian of $\slnm$ and the universal R-matrix}
\author{Adam Rej, Fabian Spill}
\date{}

\begin{document}
\begin{flushright} 
IMPERIAL-TP-AR-2010-1\\
\end{flushright}
\mbox{ }  \hfill 
\vspace{5ex}
\Large
\begin {center}     
{\bf The Yangian of $\slnm$ and the Universal R-matrix}
\end {center}
\large
\vspace{1ex}
\begin{center}
Adam Rej \ and \  Fabian Spill
\end{center}
\begin{center}
\textit{Blackett Laboratory,\\
Imperial College London\\
London SW7 2AZ, UK}\vspace{3mm}\\
\texttt{a.rej@imperial.ac.uk}\\
\texttt{fabian.spill@gmail.com}

\end{center}

\abstract{In this paper we study Yangians of $\slnm$ superalgebras. We derive the universal R-matrix and evaluate it on the fundamental representation obtaining the standard Yang R-matrix with unitary dressing factors. For $m=0$, we directly recover up to a CDD factor the well-known S-matrices for relativistic integrable models with $\sun$ symmetry. Hence, the universal R-matrix found provides an abstract plug-in formula, which leads to results obeying fundamental physical constraints: crossing symmetry, unitrarity and the Yang-Baxter equation. This implies that the Yangian double unifies all desired symmetries into one algebraic structure.  In particular, our analysis is valid in the case of $\slnn$, where one has to extend the algebra by an additional generator leading to the algebra $\glnn$. We find two-parameter families of scalar factors in this case and provide a detailed study for $\alg{gl}(1|1)$.}

\newpage

\section{Introduction}

Yangians are important algebraic structures arising in many integrable models, such as XXX spin chains, the Hubbard model \cite{Uglov:1993jy} or integrable field theories (see e.g. the reviews \cite{Bernard:1992ya}-\cite{MacKay:2004tc}). They are infinite-dimensional extensions of traditional Lie algebra symmetries and are generally associated with integrable models where the two-particle scattering matrix is a rational function depending on the difference of the particles' rapidities. More recently, it has been shown that the Yangian is also a symmetry of the planar AdS/CFT correspondence \cite{Dolan:2003uh}-\cite{Dolan:2004ps}, at least at leading order in perturbation theory and hence also of the asymptotic S-matrix \cite{Beisert:2007ds}. Interestingly, this S-matrix is not of difference form in this case, as the underlying Lie algebra is centrally extended, and the eigenvalues of the central charges are related to the rapidity $u$ in a non-trivial way. 

The mathematical reason why Yangians are related to rational S-matrices is the fact that the former possess a universal R-matrix, which can be constructed via the quantum double method. This was done explicitly for the Yangians based on simple Lie algebras in \cite{Khoroshkin:1994uk}. The Yangian is hence quasi-triangular, and the resulting R- and S-matrices on representations are thus guaranteed to satisfy the Yang-Baxter equation. In the physics literature Yangians have often been used to fix the matrix structure of R- and S-matrices. 

A distinctive feature of the Yangian is the antipode map, which corresponds to charge conjugation of the particles. This implies that the Yangian must also possess some information about the dressing factor of S-matrices on representations.

In this paper we show that, upon a carefully chosen modification of the Cartan-part of the universal R-matrix, one can recover the complete S-matrix of the $\alg{su}(n)\times \alg{su}(n)$ Principal Chiral Field \cite{Berg:1977dp,Ogievetsky:1987vv}, or, up to a CDD factor, S-matrices of other $\alg{su}(n)$ integrable models, such as the $\sun$ Gross-Neveu model. We evaluate the whole matrix structure of the universal R-matrix recovering the expected Yang R-matrix, which is proportional to $R=1+\frac{1}{u}\perm$, with $\perm$ being the permutation operator. Such explicit evaluation can be seen as a test on the ordering of the root part of the universal R-matrix, which, to our knowledge, has not been proved rigorously yet. The advantage of our approach is that we provide a plug-in formula for the universal R-matrix, so we do not need to explicitly solve the Yang-Baxter equation and the crossing equation. Indeed, quasi-triangularity of the Double Yangian guarantees these to be satisfied. We note that the universal R-matrices of \cite{Khoroshkin:1994uk} do not lead to unitary R-matrices, but still capture the essential Gamma functions of the scalar factor. This is related to the fact that different, inequivalent prescriptions used for diagonalisation of the Cartan part lead to quasi-triangular R-matrices, and only some of them respect unitarity.

Our analysis remains valid for all Yangians based on the simple Lie superalgebras $\slnm$, and the results for $\alg{su}(n)$ are merely a special case. We also study the non-simple $\slnn$ algebras, which are interesting from the physics point of view. The quantum double and the universal R-matrix require an extension of the algebra to $\glnn$. In this way we generalise the results of \cite{Spill:2008yr} for $\gltt$ which are of relevance for the AdS/CFT correspondence. 

On the fundamental representation we find that the R-matrix is, as expected, proportional to the generalisation of Yang's R-matrix, 
\beq
R = \bar{R}_0(u) \left(\frac{u}{u+1} + \frac{1}{u+1}\perm \right)\,,
\eeq
where $\perm$ is now the graded permutation operator. The non-unitary dressing factor for $n \neq m$ is very similar to that of $\alg{sl}(n)$
\beq
\bar{R}_0 (u)=\frac{\Gamma \left(\frac{1-u}{n-m}\right) \Gamma
   \left(\frac{u}{n-m}\right)}{\Gamma
   \left(-\frac{u}{n-m}\right) \Gamma
   \left(\frac{u+1}{n-m}\right)}, \quad n\neq m\,.
\eeq
It depends on $n$ and $m$ only via the difference $n-m$, which coincides with the dual Coxeter number of $\slnm$. Hence, we find that the dressing factor of $\slnm$ is the same as the one of $\alg{sl}(n-m)$. Interestingly, this formula is also valid in the case $n = m\pm 1$, when the ratio of the Gamma functions actually degenerates to a rational function. We note that the Yangian and universal R-matrix in the case $\slnm$, $n\neq m$ have been discussed in \cite{Stukopin:2005aa}. However, we could not find in this paper several details necessary for the evaluation of the universal R-matrix on representations, such as normalisation constants, shifts in the spectral parameter as well as the conventions regarding fermionic generators. Furthermore, it seems necessary to correct some of the Serre relations, see \cite{Gow:2007th}.

For $n=m$, there is some freedom in the definition of $\mathcal{Y}(\glnn)$ since $\glnn$ possess a one-dimensional centre as well as a one-dimensional external automorphism. In particular, one can rescale the automorphism and add a multiple of the central element to it without modifying the commutation relations. Even though all such choices of rescaling and shift give isomorphic Yangians, they lead to different dressing factors of the R-matrix on representations. Note that this case shows the power of the universal R-matrix approach, as we can calculate R-matrices with their dressing factors without direct reference to any underlying crossing equation. Indeed, the antipode acts on the simple components of the $\glnn$ Yangian trivially and superficially one is lacking a precise derivation of a crossing equation. However, the antipode acts non-trivially on the outer automorphism of $\glnn$ and shifts it by $\mu n$, where $\mu$ is the parameter related to the rescaling of the automorphism. Hence, we conjecture the generalised crossing equation for $\glnn$ to be of the form $S(u) S(\mu n -u) = h(u,\mu)$, where $h(u,\mu) $  is a rational function. We have found evidence that this conjecture is accurate in the simplest case of $\alg{gl}(1|1)$. Physical models with $\slnn$ symmetry are often special and the complicated dressing factors found might lead to interesting and rich physics.

Our paper is organised as follows. We begin by recalling the definition of the $\slnm$ Lie algebras in section \ref{sec:LSA} as well as the necessary extension to $\glnn$ in the case $n=m$. In section \ref{sec:yangian} we define the Yangian of $\slnm$ and $\glnn$. We will mostly use Drinfeld's second realisation, which is suitable for the construction of the universal R-matrix as done in section \ref{sec:yangdoub}. Finally, we evaluate the universal R-matrix in section \ref{sec:funR}. We delegate the technical discussion of the q-deformed Cartan matrices and their inverses to Appendix \ref{sec:Aanditsinverse}.  In Appendix \ref{sec:diff} we evaluate the R-matrix for some of the non-distinguished Dynkin diagrams and conjecture that the R-matrix is unaffected by the choice of the simple roots.

\section{$\slnm$ Lie superalgebras} \label{sec:LSA}

\subsection{The special linear superalgebra $\slnm$}\label{sec:slnm}

Let us start by giving the 
definition of the Lie superalgebra $\alg{sl}(n|m)$. Its distinguished Dynkin diagram is presented in Figure \ref{Dynkindiagram}. The corresponding symmetric Cartan matrix is given by

\begin{figure}\centering
\setlength{\unitlength}{1pt}%
\small\thicklines%
\begin{picture}(260,20)(-10,-10)
\put(  5,00){\circle{15}}%
\put(  3.5,12){1}%
\put(  12,00){\line(1,0){21}}%
\put( 40,00){\ldots}%
\put( 57,00){\line(1,0){21}}%
\put( 85,00){\circle{15}}%
\put( 80.5,12){n-1}%
\put( 92,00){\line(1,0){21}}%
\put(120,00){\circle{15}}%
\put(117.5,12){n}%
\put(127,00){\line(1,0){21}}%
\put(155,00){\circle{15}}%
\put(149,12){n+1}%
\put(162,00){\line(1,0){21}}%
\put(190,00){\ldots}%
\put(207,00){\line(1,0){21}}%
\put(235,00){\circle{15}}%
\put(226,12){n+m-1}%
\put( 115,-5){\line(1, 1){10}}%
\put( 115, 5){\line(1,-1){10}}%
\end{picture}
\caption{The distinguished Dynkin diagram of $\slnm$.}\label{Dynkindiagram}
\end{figure}

\[\label{def:cartnm}
\cartnm = \begin{pmatrix}
2 &-1&0&\dots & &  & & & & \\
-1&2&-1&\dots &\vdots & & &0 & & \\
0&\dots & \ddots & -1&0& & & & & \\
\vdots&\dots&-1&2 &-1 & & & & &\\
0&\dots& 0&-1& 0& 1 & 0 &\dots  & & \\
& & & &1 &-2 &1&0&\dots&\\
& & & &0 &1&-2&1&\dots&\\
& & 0& &\vdots & & & \ddots&1&\\
& & & &0 &\dots& &1&-2
\end{pmatrix} .
\]
The symmetric Cartan matrix can also be written as $(\cartnm)_{i,j} = (\alpha_i,\alpha_j)$, i.e. it describes the scalar product of the simple roots $\alpha_i$. It is related to the usual, unsymmetric Cartan matrix $\carttnm$ (see e.g. \cite{Frappat:1996pb})
by $(\carttnm)_{i,j}=(\Delta \cartnm)_{i,j}$, where $\Delta$ is a diagonal Matrix with the first $n$ diagonal
 entries being $1$ and the last $m-1$ being equal to $-1$. 

The corresponding Chevalley-Serre generators $\gen{H}_i, \gen{E}^\pm_i$, $i=1,\dots, n+m-1$, satisfy the usual commutation relations
\<
\comm{\gen{H}_i}{\gen{H}_j} &=& 0, \nonumber\\
\comm{\gen{H}_i}{\gen{E}^\pm_j} &=& \pm {\cartnm_{ij}}\gen{E}_j^\pm,\nonumber  \\
\scomm{\gen{E}^+_i}{\gen{E}^-_j} &=& \delta_{ij}{\gen{H}_i}\,,
\>
and the following further identities
\<\label{eq:serre}
\adj{{\gen{E}}_i^\pm}{1+|(\cartnm)_{ij}|}{{\gen{E}}_j^\pm}= 0, \nonumber\\
\acomm{\comm{{\gen{E}}_	n^\pm}{\gen{E}_{n-1}^\pm}}{\comm{{\gen{E}}_n^\pm}{\gen{E}_{n+1}^\pm}} = 0 .
\>
Whereas the first identity has the same structure as the usual Serre relation for simple Lie algebras, the second one is particular for Lie superalgebras of type $\slnm$, see e.g. \cite{Grozman:1997ms}. For simplicity, we will refer to relations (\ref{eq:serre}) as Serre relations. In the distinguished basis  $\gen{E}_{n}^\pm$ are odd (fermionic) generators, while all other generators are even (bosonic). We will use the following notation
\beqa \nonumber
\adj{\gen{X}}{}{\gen{Y}} &:=& \scomm{\gen{X}}{\gen{Y}}:= \gen{X}\gen{Y} - (-1)^{|\gen{X}||\gen{Y}|}\gen{Y}\gen{X}\,, \\ \nonumber
\comm{\gen{X}}{\gen{Y}} &:=& \gen{X}\gen{Y} - \gen{Y}\gen{X}\,,\\  \nonumber
\acomm{\gen{X}}{\gen{Y}} &:=& \gen{X}\gen{Y} + \gen{Y}\gen{X}\,, \\
\sacomm{\gen{X}}{\gen{Y}} &:=&  \gen{X}\gen{Y} + (-1)^{|\gen{X}||\gen{Y}|}\gen{Y}\gen{X}\,.
\eeqa
In the case of $\slnn$ the Cartan matrix is degenerate and the algebra is not simple. Indeed, the Cartan generator
\[
\gen{C}:= \gen{H}_1 + 2 \gen{H}_2 + \dots +(n-1)\gen{H}_{n-1}+ n\gen{H}_n + (n-1)\gen{H}_{n+1} + \dots + \gen{H}_{2n-1}
\]
corresponding to the zero eigenvalue of the Cartan matrix
is central. To obtain a simple Lie algebra $\alg{psl}(n|n)$ one needs to divide out this one dimensional centre. However, $\alg{psl}(n|n)$ does not have the fundamental matrix representation required by physical applications. Furthermore, it does not possess a non-degenerate invariant bilinear form. Since the corresponding derived algebra is again $\alg{sl}(n|n)$, the central element has vanishing inner product with any other generator. To overcome this obstacle we need to extend the algebra by its external automorphism $\gen{H}_{2n}$, which 
acts on the remaining Chevalley generators as follows:
\<
\comm{\gen{H}_{2n}}{\gen{H}_j} &=& 0, \nonumber\\
\comm{\gen{H}_{2n}}{\gen{E}^\pm_n} &=& \pm \mu\gen{E}_n^\pm,\nonumber  \\
\comm{\gen{H}_{2n}}{\gen{E}^\pm_j} &=& 0,\quad j\neq n .
\>
This results in the extended non-degenerate symmetric Cartan matrix
\beqa\label{def:cartnn}
\nonumber && \cartnn = \begin{pmatrix}
2 &-1&0&\dots & \dots &\dots  & \dots & \dots & \dots &0 \\
-1&2&-1&0 &\dots & \dots & \dots & \dots & \dots & \vdots\\
0&\ddots & \ddots & -1& 0& \dots & \dots & \dots & \dots &\vdots \\
\vdots&\dots&-1&2 &-1 & \ddots & \dots &\dots &\dots & 0 \\
0&\dots& 0&-1&0&1&0&\dots  &\dots &\mu \\ \vdots
& \dots & \dots & \dots &1 &-2 &1&0&\dots&0 \\ \vdots
& \dots & \dots & \dots &0 &1&-2&1&\ddots&\vdots\\ \vdots
& \dots & \dots & \dots &\vdots & \dots & \ddots& \ddots&1&\vdots\\ \vdots
& \dots & \dots & \dots &0 &\dots& \dots &1&-2&0
\\ 0
&\dots &\dots&0 &\mu&0&\dots &\dots&0&\lambda\mu
\end{pmatrix} 
.\\
\eeqa
In principal, $\lambda$ and $\mu$ can be arbitrary complex numbers. This is related to the fact that $\slnn$ is an ideal in $\glnn$, i.e. the generator $\gen{H}_{2n}$ does not appear on the right-hand side of commutators of $\slnn$. Hence, one can rescale or add a multiple of the central element to $\gen{H}_{2n}$ without qualitatively changing the commutation relations. A canonical choice for these constants is given by $\lambda = 0$ and $\mu=1$. The
 resulting algebra is the algebra of all $2n\times 2n$ supermatrices, $\glnn$. In this paper we will always work with $\glnn$ in the case $n=m$.

\subsection{Fundamental Representation}\label{sec:funrep}

Let us briefly discuss the well-known fundamental representation for the generators of $\glnm$. We consider a graded $n+m$ dimensional vector space. The even and odd subspaces are spanned by the vectors $V_j$ with $j = 1,\dots,n$ and $j= n+1,\dots,n+m$ respectively.
The standard realisation of these is simply the vector $V_j$ with the element $1$ in the $j$-th row and $0$ otherwise. Let $E_{ij}$ 
denote matrices with entry $1$ for the element $(i,j)$ and $0$ otherwise, i.e. they act as
\[
E_{ij}V_k = \delta_{jk}V_i .
\]
Then, on the fundamental
representation of $\slnm$ the Chevalley-Serre basis is realised as follows
\beqa
\csgen{H}_i &=& E_{i,i}-E_{i+1,i+1},\quad i<n\,, \nonumber\\
\csgen{H}_n &=& E_{n,n}+E_{n+1,n+1},\quad \nonumber\\
\csgen{H}_i &=& E_{i+1,i+1}-E_{i,i},\quad n<i<n+m\,, \nonumber\\
\csgen{E}^+_i &=& E_{i,i+1},\nonumber\\
\csgen{E}^-_i &=& E_{i+1,i},\quad i<n\,, \nonumber\\
\csgen{E}^-_i &=& -E_{i+1,i},\quad n\leq i<n+m\,.
\eeqa
The additional Cartan generator for $\alg{gl}(n|n)$ is represented by
\[
\csgen{H}_{2n} = \frac{\mu}{2}\left(\sum_{i=1}^{n}E_{i,i} - \sum_{i=n+1}^{2n}E_{i,i}\right) + \frac{\lambda}{2 n}\sum_{i=1}^{2n}E_{i,i}\,.
\]

\section{The Yangian of $\slnm$ and $\glnn$}\label{sec:yangian}

The Yangian in Drinfeld's second realisation for $\slnm$ is defined by the generators 
$\csgen{H}_{i,k},\csgen{E}^\pm_{i,k}$, with $k=0,1,\dots$ and $i=1,\dots n+m-1$. These generators satisfy the commutation relations \cite{Gow:2007th}-\cite{Stukopin:2005aa}
\begin{align}
\label{def:yangian2}
&[\csgen{H}_{i,k},\csgen{H}_{j,l}]=0,\quad [\csgen{H}_{i,0},\csgen{E}^+_{j,k}]=A_{ij} \,\csgen{E}^+_{j,k},\nonumber\\
&[\csgen{H}_{i,0},\csgen{E}^-_{j,k}]=- A_{ij} \,\csgen{E}^-_{j,k},\quad \scomm{\csgen{E}^+_{i,k}}{\csgen{E}^-_{j,l}}=\delta_{i,j}\, \csgen{H}_{j,k+l},\nonumber\\
&[\csgen{H}_{i,k+1},\csgen{E}^\pm_{j,l}]-[\csgen{H}_{i,k},\csgen{E}^\pm_{j,l+1}] = \pm\frac{1}{2} A_{ij} \{\csgen{H}_{i,k},\csgen{E}^\pm_{j,l}\},\nonumber\\
&\scomm{\csgen{E}^\pm_{i,k+1}}{\csgen{E}^\pm_{j,l}}-\scomm{\csgen{E}^\pm_{i,k}}{\csgen{E}^\pm_{j,l+1}} = \pm\frac{1}{2} A_{ij} \sacomm{\csgen{E}^\pm_{i,k}}{\csgen{E}^\pm_{j,l}},
\end{align}
\begin{eqnarray} 
&&Sym_{\{k\}}[\csgen{E}^\pm_{i,k_1},[\csgen{E}^\pm_{i,k_2},\dots [\csgen{E}^\pm_{i,k_{1+|A_{ij}|}}, \csgen{E}^\pm_{j,l}\}\dots\}\}=0,\nonumber\\
&&\acomm{\comm{{\csgen{E}}_{n,0}^\pm}{\csgen{E}_{n-1,k}^\pm}}{\comm{\csgen{E}_{n,0}^\pm}{\csgen{E}_{n+1,l}^\pm}} = 0\,.
\end{eqnarray}
In the case of $\glnn$ there exist a set of additional generators $\csgen{H}_{2n,k}$, which satisfy the above relations when extended to $i=2n$.

The fundamental evaluation representation of the Yangian generators with a spectral parameter $u$ is given by

\beqa
\csgen{H}_{i,k} &=& (u+a_i)^k\csgen{H}_i\,, \nonumber\\
\csgen{E}^+_{i,k} &=& (u+a_i)^k\csgen{E}^+_{i}\,,\nonumber\\
\csgen{E}^-_{i,k} &=& (u+a_i)^k\csgen{E}^-_{i}\,.
\eeqa
The shift parameter $a_i$ may be directly determined from (\ref{def:yangian2}). One finds

\[
 a_i = \left\{\begin{tabular}{cc}
             $\frac{i}{2}$,&\quad $i\leq n$,\\
	     $\frac{2n-i}{2}$,&\quad $n<i<n+m$ . \\
             \end{tabular}\right.
\]
In the case $n=m$ the additional Cartan generators $\csgen{H}_{2n}$ are represented by 
\[\label{def:h2nyang}
\csgen{H}_{2n}=(u+a_n + \frac{\lambda}{2n})^k\csgen{H}_{2n}\,. 
\]

\subsection{Root ordering}\label{sec:rootorder}

Since the universal R-matrix contains products of infinitely many non-commutative generators, the ordering of those generators is of crucial importance. We will use the general prescription given in \cite{Khoroshkin:1994uk}, which is as follows. If two positive roots of the Yangian $\gamma_1$, $\gamma_2$ have already been ordered and $\gamma_1 + \gamma_2 = \gamma_3$, then, if $\gamma_1<\gamma_2$, one puts $\gamma_1<\gamma_3<\gamma_2$. As the Yangian is a deformation of the polynomial algebra, one can write $\gamma = \alpha + n\delta$. Here, $\alpha$ is a positive root of the algebra $\alg{g}$ and $\delta$ is the imaginary root. For the proper Yangian $n>0$, while $n\in\mathbb Z$ for the Yangian double. The associated positive root generator is denoted by
\[
 \cwgen{E}^+_{\gamma}= \cwgen{E}^+_{\alpha+n\delta} = \cwgen{E}^+_{\alpha,n}\,.
\]
The corresponding negative root generator is
\[
 \cwgen{E}^-_{\gamma}= \cwgen{E}^-_{\alpha+n\delta} = \cwgen{E}^-_{\alpha,n}.
\]
In the case of $\mathcal{Y}(\glnm)$ we will first order the roots of $\glnm$. If $\alpha_1, \dots, \alpha_{n+m-1}$ denote 
the simple roots, then we define the set of all positive roots as follows. Let $\beta_{[k,l]}$, with $k=1,\dots, n+m-1$ and $0\leq l<k$, be one of the positive roots labelled by a double index $[k,l]$. In that case
\beqa
\beta_{[k,0]} &=& \alpha_k\,, \nonumber\\
\beta_{[k,l]} &=& \alpha_k+\alpha_{k-1}+\dots,\alpha_{k-l} .
\eeqa 
Now, we order the roots such that
\[
\beta_{[k_1,l_2]}<\beta_{[k_2,l_2]}\,,
\]
if $k_1<k_2$, or $k_1=k_2$ and $l_1<l_2$. One can easily check that the above definition of the root ordering is satisfied. From now on we will assume this root ordering and only occasionally use the double index notation explicitly. 

Let us now discuss the ordering of the Yangian. If $\gamma_1 = \alpha_1 + n_1\delta$ and $\gamma_2 = \alpha_2 + n_2\delta$, then $\gamma_1<\gamma_2$ if $\alpha_1<\alpha_2$, or if $\alpha_1=\alpha_2$ and $n_1<n_2$. We did not prove this ordering for Yangian in full generality, but it certainly holds on the fundamental evaluation representation.

\section{The Yangian Double and the Universal R-matrix}\label{sec:yangdoub}

The Yangian Double $\DY{\alg{g}}$ is the Drinfeld Double of the Yangian $\Ya{\alg{g}}$, i.e. the tensor product of 
$\Ya{\alg{g}}$ with its dual vector space $\Ya{\alg{g}}^*$ equipped with a product and coproduct dual to the coproduct and product of $\Ya{\alg{g}}$. A Chevalley-Serre type basis for $\Ya{\slnm}^*$ is given by the generators 
$\csgen{H}_{i,k},\csgen{E}^\pm_{i,k}$, with $k=-1,-2,\dots$ and $i=1,\dots, n+m-1$ and with an additional set of generators $\csgen{H}_{2n,k}$ in the case of $\glnn$. 
The classical analogue of the Yangian Double is the loop algebra, and a generator with the index $k$ can just be thought of as $\gen{X}_k = u^k \gen{X}_0$. However this basis is not a dual basis for the Yangian. The inner product is given by

\<
\sprod{\csgen{E}^+_{i,k}}{\csgen{E}^-_{j,l}}=(-1)^{|i|}\sprod{\csgen{E}^-_{j,l}}{\csgen{E}^+_{i,k}}=-\delta_{ij}\delta_{k,-l-1}\,,\nonumber\\
\sprod{\csgen{H}_{i,k}}{\csgen{H}_{j,-l-1}}=-A_{ij}\,\left(\frac{A_{ij}}{2}\right)^{k-l}{k\choose l} \quad \textrm{for} \quad k\geq l ,
\>
with all other products vanishing. The inner product can be derived by requiring compatibility with the Hopf Algebra structures. In particular, the dual of the product should be identified with the coproduct and vice versa.
Here, we will not list the coproduct and the product relations for the dual generators, as we will not use them. They have the same structure as in \cite{Khoroshkin:1994uk}.

To construct the universal R-matrix, one needs to construct a dual basis with respect to the inner product. To do this, it is useful to introduce generating functions for the generators

\<
 \csgen{E}^+_i(\lambda):=\sum_{k=0}^{\infty}\csgen{E}^+_{i,k}\lambda^{-k-1}\,, \quad(\csgen{E}^+)^*_i(\lambda):=-\sum_{k=-1}^{-\infty}\csgen{E}^-_{i,k}\lambda^{-k-1}\,, \nln
\csgen{E}^-_i(\lambda):=\sum_{k=0}^{\infty}\csgen{E}^-_{i,k}\lambda^{-k-1}\,, \quad(\csgen{E}^-)^*_i(\lambda):=-\sum_{k=-1}^{-\infty}\csgen{E}^+_{i,k}\lambda^{-k-1}\,,
\>

\<
 \csgen{H}^+_i(\lambda):=1+\sum_{k=0}^{\infty}\csgen{H}_{i,k}\lambda^{-k-1}, \quad\csgen{H}^-_i(\lambda):=1-\sum_{k=-1}^{-\infty}\csgen{H}_{i,k}\lambda^{-k-1}.
\>
The parameter $\lambda$ is the formal parameter of expansion. It should be noted that on evaluation representations with spectral parameter $u$ the generating function will depend on the difference $u-\lambda$, so effectively $\lambda$  may be interpreted as the spectral parameter. 

The dual of the function $\gen{J}(\lambda_1) = \sum_{k=0}^{\infty}\gen{J}_k \lambda^{-k-1}_1$ is defined as the function $\gen{J}^*(\lambda_2)= -\sum_{k=-1}^{-\infty}\gen{J^*}_k \lambda^{-k-1}_2$ such that
\<
\sprod{\gen{J}(\lambda_1)}{\gen{J}^*(\lambda_2)} = \frac{1}{\lambda_1-\lambda_2}\,.
\>
This is equivalent to introducing the generator $\gen{J}^*_l$ dual to $\gen{J}_k$ in the sense of

\[
\sprod{\gen{J}_k}{\gen{J}^*_{-l-1}}=-\delta_{k,l}.
\]
According to this definition the root generators are already written in terms of a dual basis. What remains to be found is the dual basis for the Cartan generators. Note that here the superscripts $\pm$  indicate the expansion of $\csgen{H}^\pm_i(\lambda)$ at $\lambda=0$ and $\lambda=\infty$ respectively. On evaluation representations one finds that $\csgen{H}^+_i(\lambda)$ and $\csgen{H}^-_i(\lambda)$ represent formally the same function. Their scalar product is given by \cite{Khoroshkin:1994uk}
\<
 \sprod{\gfhp{i}{1}}{\gfhm{j}{2}} = \frac{\lambda_1-\lambda_2 + \frac{A_{ij}}{2}}{\lambda_1-\lambda_2 -\frac{A_{ij}}{2}}.
\>
It turns out to be useful to consider the formal logarithms $\log(\csgen{H}^\pm_i(\lambda))$ due to the following property 

\<
\sprod{\log(\gfhp{i}{1})}{\log(\gfhm{j}{2})} = \log\frac{\lambda_1-\lambda_2 + \frac{A_{ij}}{2}}{\lambda_1-\lambda_2 -\frac{A_{ij}}{2}}\,.
\>
Therefore, 

\<
\sprod{\frac{d}{d\lambda_1}\log(\gfhp{i}{1})}{\log(\gfhm{j}{2})} = \frac{1}{\lambda_1-\lambda_2 +\frac{A_{ij}}{2}}-\frac{1}{\lambda_1-\lambda_2 -\frac{A_{ij}}{2}}\,.
\>
If one introduces the shift operator

\[
 T f(\lambda_2)= f(\lambda_2+1),
\]
then the above formula may be written as
\[\label{eq:diagform}
\sprod{\frac{d}{d\lambda_1}\log(\gfhp{i}{1})}{\log(\gfhm{j}{2})} = (T^{-A_{jk}/2}-T^{A_{jk}/2})\frac{\delta_{ik}}{\lambda_1-\lambda_2}\,.
\]
This is a matrix equation and to complete the task of the diagonalisation one needs to invert the 
operator 
\[\label{def:dop}
D_{ij} = T^{-A_{ij}/2}-T^{A_{ij}/2}\,.
\]
Note that on evaluation representations $T$ effectively shifts the spectral parameter $u_2$ due to the aforementioned fact that the Drinfeld currents depend on the difference $\lambda-u$. For the sake of the following discussion it is useful to introduce the q-deformed symmetric Cartan matrix, i.e. we replace each number $x$ by its q-number, see Appendix \ref{sec:Aanditsinverse} for further details. Explicitly, 
\beq
x \to [x]_q = \frac{q^x-q^{-x}}{q-q^{-1}}\,,
\eeq
so the q-deformed Cartan matrix takes the following form 
\[ \label{eq:Aqdeformiert}
 A(q)_{ij} = \qnumb{(\alpha_i,\alpha_j)} = \qnum{(\alpha_i,\alpha_j)}.
\]
The $D_{ij}$ operator is then related to the q-deformed Cartan matrix through
\[\label{eq:dascartan}
D_{ij} = -(T^{1/2} - T^{-1/2})A(T^{1/2})_{ij}\,.
 \]
The q-deformed Cartan matrices, their inverses and determinants are discussed in Appendix \ref{sec:Aanditsinverse}. 

\subsection{The Universal R-Matrix}

In the previous section we have established the Yangian double $\DY{\alg{\slnm}}$ by
generalising the analysis of \cite{Khoroshkin:1994uk} in the case of simple Lie algebras. The universal R-matrix can now be easily stated with the help of the diagonalised form \eqref{eq:diagform} since it is simply the canonical element of the Yangian double, i.e. the sum over all elements of the Yangian $\mathcal{Y}({\alg{\slnm}})$ tensor its appropriate dual. One should stress that the Yangian consists not only of the Chevalley-Serre generators \eqref{def:yangian2} and their commutators, but also of all powers of the corresponding generators. Schematically, the dual product decomposes as follows
\<
\sprod{E^+ H E^- }{(E^+)^{*}(H)^{*}(E^-)^{*}} = \sprod{E^+}{(E^+)^*}\sprod{H}{H^*}\sprod{E^-}{(E^-)*},
\>
just as for simple Lie algebras \cite{Khoroshkin:1994uk}. Hence, the universal R-matrix has the quasi-triangular structure
\beq\label{eq:rsplit}
R = R_E R_H R_F\,.
\eeq
The positive and negative root parts are given in terms of ordered products
\beqa\label{def:Rpm}
R_E &=& \prod_{\beta,k\geq 0}^\rightarrow \exp(-(-1)^{|\beta|}\mathcal{F}^{|\gamma|}\cwgp{\beta+k\delta}\otimes \cwgm{\beta-(k+1)\delta})\,,\nln
R_F &=& \prod_{\beta,k\geq 0}^\leftarrow \exp(-\mathcal{F}^{|\beta|}\cwgm{\beta+k\delta}\otimes \cwgp{\beta-(k+1)\delta}).
\eeqa
Here, $\mathcal{F}$ is the usual Fermi-number generator.  An important feature of (\ref{def:Rpm}) is that the product in $R_E$ is taken in the order specified in section \ref{sec:rootorder}, whereas for $R_F$ the reverse ordering is applied. The product is only taken over positive roots $\beta \in \slnm$ and the symbol $\delta$ denotes the imaginary root. The Cartan part of the universal R-matrix is significantly more complicated
\[\label{def:RH}
R_H = \prod_{i,j}\exp\left(\sum_{t=0}^\infty\left(\left(\frac{d}{d\lambda_1}\log(\gfhp{i}{1})\right)_{t}\otimes\left( D^{-1}_{ij}\log(\csgh{j}^-(\lambda_2)\right)_{-(t+1)}\right)\right)\,.
\]
Here, the superscripts $t$ and $-(t+1)$ denote the respective coefficients of the expansion of the generating functions in $\lambda_1 \gg 1$, $\lambda_2 \ll 1$. Tensoring them together is thus equivalent to taking the residue at $\lambda_1 = \lambda_2$.  In the case of integer-valued Cartan matrices we find that the inverse of the q-Cartan matrix is also q-integer valued up to an overall constant q-number $\qnumb{l}$.

The concrete definition of the inverse $D^{-1}_{ij}$ on a set of functions of spectral parameter determines the scalar part of the universal matrix. According to \eqref{eq:dascartan}, the operator $D$ may be expressed solely through the translation operator $T$, for which the action on functions of spectral parameter is well-definied. Clearly, one may define $D^{-1}_{ij}$ by expanding it in a power series around either $T=0$ or $T=\infty$. It turns out, however, that the both expansions \textit{do not} result in the same scalar part $R_H$. We would like to argue that the guiding principle should be unitarity. Indeed, the Yangian Double is not triangular, i.e. the equation 
\[
R_{12} R_{21} = \idm
\]
does not hold. Only a balanced expansion in power series in $T$ and $T^{-1}$ will lead to unitary dressing factors. It follows immediately from \eqref{eq:Aqdeformiert} that $A(q^{-1}) = A(q)$ so that the operator $D_{ij}$ defined in \eqref{eq:dascartan} satisfies $D_{ij}(q) = - D_{ij}(q^{-1})$. The same must hold for its inverse $D^{-1}_{ij} (q)$ thus one may write
\beq \label{eq:Dunitary}
D^{-1}_{ij} (q) = \tfrac{1}{2} \left(D^{-1}_{ij}(q)-D^{-1}_{ij}(q^{-1}) \right)\,.
\eeq
We propose to expand the first term at $q=0$ and the second one at $q=\infty$ setting  afterwards $q \to T^{1/2}$ in the expansions. Subsequently, the principle branch of the square root should be applied.  We conjecture that the resulting R-matrix contains a unitary dressing factor satisfying the corresponding crossing equation. Moreover, in all cases studied in what follows the dressing factor found in this way is a \textit{meromorphic} function up to a square root of a CDD factor. This suggest that this may be a general feature of this procedure. Please note also that analytic properties   of a given solution to crossing and unitarity equations are dictated by the concrete physical model and cannot be determined with help of the universal R-matrix.

Generic Cartan matrices of superalgebras, in particular those corresponding to the $\glnn$ algebra with arbitrary extension parameters $\lambda, \mu$ , have non-integer elements and the aforementioned prescription needs to be applied. Integer-valued Cartan matrices allow for further simplification since there exists a matrix $C(q)$
\beq
 A(q) C(q) = \qnumb{l} \textit{Id}\,,
\eeq
such that its elements are \text{polynomials} in $q$ and $q^{-1}$. Here, $l$ is assumed to take minimal value for which such $C(q)$ exists.
The inverse of \eqref{eq:dascartan} may now be written as
\beqa\label{eq:dinv}
D^{-1} =C(T^{1/2})\frac{1}{T^{-l/2}-T^{l/2}}\,.
\eeqa
Expanding the second term in the vicinity of $T=(\infty)^{\mp 1}$ gives
\[\label{RHpos}
R^{\pm}_H = \prod_{i,j,k,t}\exp\left(\pm\left(\frac{d \log(\gfhp{i}{1})}{d\lambda_1}\right)_{t}\otimes\left( C_{ij}(T^{1/2})\log(\csgh{j}^-(\lambda_2\pm(k+1/2)l)\right)_{-t-1}\right)\,.
\]
For $\sln$ algebras $l=n$ and the above formulae reduce to the one proposed in \cite{Khoroshkin:1994uk}. The scalar part leading to a unitary dressing factor may be formally written as
\beq
R_{H} = \sqrt{ \frac{R^{+}_H}{R^{-}_H}}\,.
\eeq
This formula, however, remains also valid for the supersymmetric counterpart $\slnm$ with $l = n-m$ and $n \neq m$. Clearly, the case of $n=m$ is special and the definition of $C(q)$ becomes ambiguous for non-canonical choices of the extension parameters. Thus the matrix $C(q)$ is convenient for classification purposes, but becomes ill-defined in the general case of real-valued Cartan matrices.

\section{The R-matrix on the fundamental representation}\label{sec:funR}

In this section we will systematically evaluate the universal R-matrix on the 
fundamental representation of $\glnm$ and its Yangian, which we introduced in section \ref{sec:funrep}.

The first step towards the explicit evaluation of the Cartan part \eqref{def:RH} of the universal R-matrix is to invert the operator $D$ given in (\ref{def:dop}). In Appendix \ref{sec:Aanditsinverse} we report on the generic formula of the q-deformed inverse Cartan matrices for $\slnm$, $n \neq m$ as well as for $\glnn$ with arbitrary parameters $\lambda$ and $\mu$. This leads immediately to $D^{-1}$ via (\ref{eq:Dunitary}). These results are crucial to perform the second step.

\subsection{Evaluating the universal R-matrix}

In this section we will evaluate the R-matrix \eqref{eq:rsplit} on the fundamental representation of $\slnm$. The factors $R_E$ and $R_F$ are fairly easy to evaluate due to the nilpotence of the roots
\beqa \label{REfun}
R_E&=&\prod^{\frac{1}{2} (m+n-1) (m+n)}_{k=1} \mbox{exp}(\frac{1}{u}(-1)^{|\beta_k|}\mathcal{F}^{|\beta_k|}E^{+}_{\beta_k}\otimes E^{-}_{\beta_k})\nln
&=&\prod^{\frac{1}{2} (m+n-1) (m+n)}_{k=1} \left(1+\frac{1}{u}(-1)^{|\beta_k|}\mathcal{F}^{|\beta_k|}E^{+}_{\beta_k}\otimes E^{-}_{\beta_k}\right) \,,
\eeqa
\beqa \label{RFfun}
R_F&=&\prod^{\frac{1}{2} (m+n-1) (m+n)}_{k=1} \mbox{exp}(\frac{1}{u}\mathcal{F}^{|\beta_k|}E^{-}_{\beta_k}\otimes E^{+}_{\beta_k})\nln
&=&\prod^{\frac{1}{2} (m+n-1) (m+n)}_{k=1} \left(1+\frac{1}{u}\mathcal{F}^{|\beta_k|}E^{-}_{\beta_k}\otimes E^{+}_{\beta_k}\right)\,.
\eeqa
Here, $u=u_1-u_2$ is the difference of the spectral parameters of both factors of the tensor product. The order in the products $R_E$ and $R_F$ is taken as outlined in section \ref{sec:rootorder}. On the fundamental representation the individual blocks commute, however. Thus what remains is the evaluation of \eqref{def:RH}.  Let us first show that the factor \eqref{def:RH} is convergent. Indeed, each element of $R_H=\prod_{ij} (R_H)_{ij}$ is of the form

\[
(R_H)_{ij} \sim \prod_{n=0}^{\infty} \frac{a\,n + b}{a\,n + b + h_i}\frac{a\,n + b+h_i - h_j}{a\,n + b - h_j}\,.
\]
Using the following product representation of the Gamma function
\beq
 \Gamma(z) = \lim_{M\rightarrow\infty}\frac{1}{z}e^{-z(\sum_{k=1}^M 1/k - \log M)} \prod_{n=1}^{M} \frac{1}{1+z/n}e^{z/n}\,,
\eeq
one easily finds
\beq
(R_H)_{ij} \sim  \prod_{n=0}^{\infty} \frac{a\,n + b}{a\,n + b + h_i}\frac{a\,n + b+h_i - h_j}{a\,n + b - h_j}= \frac{\Gamma(\frac{b+h_i}{a})\Gamma(\frac{b- h_j}{a})}
{\Gamma(\frac{b+h_i - h_j}{a})\Gamma(\frac{b }{a})}\,.
\eeq
The matrix \eqref{def:RH} is diagonal since the Cartan algebra elements are diagonal. Using the prescription \eqref{eq:Dunitary} one finds
\beq \label{R0mix}
(R_H)_{11; 11}\equiv R_0 (u)= \left\{ \begin{array}{ccc}\sqrt{h(u)} \frac{ \Gamma \left(\frac{1-u}{n-m}\right) \Gamma
   \left(\frac{u}{n-m}\right)}{\Gamma
   \left(-\frac{u}{n-m}\right) \Gamma
   \left(\frac{u+1}{n-m}\right)}, \quad n \neq m, \\ \\
\frac{u+\frac{1}{2}}{u-\frac{1}{2}}, \quad n=m.  \end{array} \right.
\eeq
The function $h(u)$ is a simple ratio of trigonometric functions and, as follows from the discussion in the subsequent section, may be dropped being solely a CDD factor. Surprisingly, \eqref{R0mix} coincides for $m=0$ with the $u(n)$ dressing factor found in \cite{Berg:1977dp}. Moreover, for $m >0$ and $m \neq n$ it is identical to the $u(N)$ dressing factor with $N=n-m$. 

Evaluating the remaining elements of $R_H$ and combining them with the formulas \eqref{REfun} and \eqref{RFfun} one finds the following compact result for the R-matrix
\beqa\label{eq:funR}
R &=& R_0(u) \left(\frac{u}{u+1} + \frac{1}{u+1}\perm \right)\,.\nonumber\\
\eeqa
Here, $\perm$ denotes the graded permutation operator
\[
\perm V_i\otimes V_j = (-1)^{|i||j|}V_j\otimes V_i.
\]
The matrix part of this R-matrix is a supersymmetric version of Yang's R-matrix. It does not depend on whether the expansion in the Cartan part is taken for $T \ll1$ or $T \gg 1$. This is expected, as this is the only solution to the rational Yang-Baxter equation on those representation spaces.

\subsection{Crossing and unitarity} \label{sec:CandU}
The scalar factor $R_0$ is the scalar part of the R-matrix and is usually found by means of crossing and unitarity equations. Let us then inspect what effect the simplest non-unitary definition of $D^{-1}_{ij} (T)$ will have on $R_H$. More precisely, we will define the inverse as a power series at $T=0$. It is straightforward to check that now $R^{-}_0(u) R^{-}_0(-u) \neq 1$ for $n \neq m$ or $n \neq m \pm 1$\footnote{In the case of $n = m \pm 1$ the ratios of Gamma functions reduce to rational functions.}. Surprisingly, unitarity may be easily restored by rescaling with the following simple function
\beq
f(u)^{-1}=\frac{\sin{\left(\frac{\pi\,u}{n-m}\right)}}{\sin{\left(\frac{\pi\,(1 + u)}{n-m}\right)} }.
\eeq
The new scalar factor $\bar{R}_0(u)=f(u)R^{-}_0(u)$ coincides with \eqref{R0mix} up to the $\sqrt{h(u)}$ factor. This suggests that the $T=0$ expansion violates unitarity in a rather weak manner and unitary dressing factors may be obtained by rescaling with simple analytic functions.  This observation often facilitates the analytic evaluation in more complex cases.

Since the dual Coxeter number of $\slnm$ is non-zero for $m \neq n$, the usual derivation \cite{Ogievetsky:1987vv} of the crossing equation should be applicable  leading to
\beq \label{crossing}
S_0((n-m)-u)\, S_0(u)=\frac{\left(u-\frac{n-m}{2}\right)^2-\frac{1}{4} \left((n-m)-2\right)^2}{(u-(n-m))\, u}\,.
\eeq
It is easy to check that $S_0(u)=\bar{R}_0(u)$ satisfies this relation and that the aforementioned prefactor is a CDD solution
\beq
h((n-m)-u)\, h(u) =1 \,.
\eeq
For $n=m$, however, the dual Coxeter number is exactly zero. It would be interesting to derive rigorously the relativistic crossing equation in this case. The fact that the shift in the crossing equation is usually equal to the dual Coxeter number seems to be related to the action of the antipode $S$ on the Yangian generators. In the case of $n\neq m$ we find that the generators of $\mathcal{Y}(\slnm)$ satisfy 
\[
S(\genY{J}) = -\genY{J} + c\, \gen{J},
\]
with $c$ being proportional to the dual Coxeter number $l=n-m$. This result (with $c=0$) remains true for $\slnn$, so one naively does not expect any shift in the crossing equation. However, it should be noted that the automorphism \eqref{def:h2nyang} has a non-trivial antipode, which, if allowing for generic parameters $\mu$, $\lambda$, is proportional to $\mu n $ on the fundamental representation. This suggests that the overall shift is equal to $a\mu n$. The case of the $\alg{gl}(1|1)$ algebra with arbitrary $\mu$ and integer $\lambda$ studied in section \ref{sec:gl11} indicates that $a=1$. For $\lambda=0$ and $\mu=1$ the unitary scalar factor derived from the universal R-matrix does not seem to depend on $n=m$, which allows us to conjecture the following crossing equation for arbitrary $n$
\beq
\bar{R}_0(u)\, \bar{R}_0(n-u) =\frac{(u-n-\frac{1}{2})\,(u+\frac{1}{2})}{(u-n+\frac{1}{2})\,(u-\frac{1}{2})}\,.
\eeq

At this point an interesting physical aspect of (\ref{eq:rsplit}) should be stressed. Since the formula (\ref{eq:rsplit}) is a plug-in type formula, with the corresponding symmetry algebra $\mathcal{A}$ to be specified, the resulting R-matrix exhibits the very same symmetry and corresponds to the scattering of modules of $\mathcal{A}$. The ground state, i.e. the state without any physical excitations, must be a \textit{physical} vacuum and thus invariant with respect to global action of $\mathcal{A}$\footnote{If $\mathcal{A}$ is a global symmetry of the Hamiltonian.}. For example the $su(2)$ R-matrix may be used to describe the scattering of spinons on the antiferromagnetic state of the XXX spin chain, which explains the complexity of the corresponding dressing phase.

\subsection{The general solution for $\alg{gl}(1|1)$} \label{sec:gl11}
In this section we will demonstrate the efficiency of the universal R-matrix approach and derive the dressing phase of the $\alg{gl}(1|1)$ algebra for arbitrary $\mu$ and integer values of $\lambda$. Using the unitary prescription \eqref{eq:Dunitary} one finds from \eqref{eq:dascartan}

\beq
R_0(u)= \sqrt{h(u)} f(u)\, R_{0,1}(u) \, R_{0,2}(u)\,,
\eeq
with
\beq
R_{0,1}(u)=\prod^{\infty}_{n=0} \frac{\left(n-\frac{\lambda }{2}+\frac{u}{\mu
   }-\frac{1}{\mu }+1\right)_{\lambda } \left(n-\frac{\lambda }{2}+\frac{u}{\mu }+\frac{1}{\mu }+1\right)_{\lambda }}{\left(n-\frac{\lambda }{2}+\frac{u}{\mu }+1\right)^2_{\lambda }}\,,
\eeq
where $(\,.\,)_n$ denotes the Pochhammer symbol, and
\beq
R_{0,2}(u)=\frac{\Gamma \left(1+\frac{u}{\mu }\right) \Gamma \left(\frac{u}{\mu }\right) \Gamma \left(\frac{2 u-\lambda +\mu -2}{2 \mu
   }\right) \Gamma \left(\frac{2 u+\lambda +\mu +2}{2 \mu }\right)}{\Gamma \left(\frac{1+u}{\mu }\right) \Gamma
   \left(\frac{2 u-\lambda +\mu }{2 \mu }\right) \Gamma \left(\frac{u+\mu -1}{\mu }\right) \Gamma \left(\frac{2 u+\lambda
   +\mu }{2 \mu }\right)}\,.
\eeq
Interesingly, and in accordance with the discussion in section \ref{sec:CandU}, the contribution $R_{0,1}(u)\,R_{0,2}(u)$ may also be found by defining the inverse $D^{-1}$ as a power series at $T=0$. The function $f(u)$ may be thought of as the corresponding \textit{``unitarisation''} factor
\beq
f(u)=\left(\frac{\sin \left(\frac{\pi\, u}{\mu} +\frac{\pi\, \lambda}{2}\right)}{\sin \left( \frac{\pi\,(u-1)}{\mu}+\frac{\pi\, \lambda}{2}\right)}  \right)^{\lambda} \frac{\sin\left(\frac{\pi\, u}{\mu}\right)\,\cos \left(\frac{\pi\,(2u-2-\lambda)}{2\mu}\right)}{\sin\left(\frac{\pi\,(1-u)}{\mu}\right)\,\cos\left(\frac{\pi\,(2u-\lambda)}{2\mu}\right)}\,.
\eeq
With a bit of algebra one finds that $R_0(u)$ satisfies the following functional equation 
\beq \label{eq:gl11crosseq}
R_0(u)\,R_0(\mu-u)=C(u,\lambda,\mu)\,,
\eeq
with the function $C(u,\lambda,\mu)$ being different for even and odd values of lambda
\beqa
\nonumber
C(u,\lambda \in \text{even}, \mu)&=&-\frac{u (u-\mu +1) \left((\mu -2 u)^2-\lambda ^2\right)} {(u+1) (u-\mu ) (2 u+\lambda -\mu +2) (-2 u+\lambda +\mu +2)}\,\times \\
&&\frac{\left(\frac{1-u}{\mu }\right)_{\frac{\lambda }{2}+1}
   \left(\frac{u-1}{\mu }\right)_{\frac{\lambda }{2}} \left(-\frac{u+1}{\mu }\right)_{\frac{\lambda }{2}+1}
   \left(\frac{u+1}{\mu }\right)_{\frac{\lambda }{2}}}{
   \left(-\frac{u}{\mu }\right)^2_{\frac{\lambda }{2}+1}\left(\frac{u}{\mu }\right)^2_{\frac{\lambda }{2}}}\,,
\eeqa
\beqa
\nonumber
C(u,\lambda \in \text{odd}, \mu)&=&-\frac{(u-1) (u-\mu +1) \left((\mu -2 u)^2-\lambda ^2\right) }{u (u-\mu ) (2 u+\lambda -\mu +2) (-2 u+\lambda +\mu
   +2)}\,\times \\
\nonumber
&&\frac{\left(\frac{u-1+\frac{\mu}{2}}{\mu }\right)_{\frac{\lambda -1}{2}}
   \left(\frac{u+1+\frac{\mu
  }{2}}{\mu }\right)_{\frac{\lambda -1}{2}} \left(\frac{-2 u+\mu -2}{2 \mu }\right)_{\frac{\lambda
   +1}{2}} \left(\frac{-2 u+\mu +2}{2 \mu }\right)_{\frac{\lambda +1}{2}}}{\left(\frac{1}{2}-\frac{u}{\mu }\right)^2_{\frac{\lambda +1}{2}} \left(\frac{u}{\mu
   }+\frac{1}{2}\right)^2_{\frac{\lambda -1}{2}}}\,.
   \\
\eeqa
Please note that in all cases the function $C(u,\lambda,\mu)$ is rational. The prefactor $\sqrt{h(u)}$ is again a CDD solution to this equation and will be neglected. Although we do not have any further algebraic evidence, we believe \eqref{eq:gl11crosseq} to be the corresponding crossing equation for arbitrary $\mu$ and integer values of $\lambda$. The dressing factor $R_0(u)$ is substantially more complicated than any other relativistic dressing factor. This can be easily seen by piecewise analytic evaluation of $R_0(u)$. For example, for $u<2 \mu-\frac{\lambda  \mu }{2} -1$ one finds

\< \nonumber
&&\frac{R_0(u)}{f(u)}=\frac{1}{\Gamma \left(\frac{u+1}{\mu }\right) \Gamma
   \left(\frac{u+\mu -1}{\mu }\right) \Gamma \left(\frac{2 u-\lambda +\mu }{2 \mu }\right) \Gamma
   \left(\frac{2 u+\lambda +\mu }{2 \mu }\right)} e^{2 \zeta ^{(1,0)}\left(-1,\frac{u}{\mu }-\frac{\lambda }{2}+1\right)}\\ \nonumber
&&e^{-2 \zeta
   ^{(1,0)}\left(-1,\frac{u}{\mu }+\frac{\lambda }{2}+1\right)-\zeta ^{(1,0)}\left(-1,\frac{u+\mu
   -1}{\mu }-\frac{\lambda }{2}\right)+\zeta ^{(1,0)}\left(-1,\frac{\lambda }{2}+\frac{u+\mu -1}{\mu
   }\right)-\zeta ^{(1,0)}\left(-1,\frac{u+\mu +1}{\mu }-\frac{\lambda }{2}\right)}\\ \nonumber
&&e^{\zeta
   ^{(1,0)}\left(-1,\frac{\lambda }{2}+\frac{u+\mu +1}{\mu }\right)} \Gamma \left(\frac{u}{\mu
   }-\frac{\lambda }{2}+1\right)^{\lambda -\frac{2 u}{\mu }} \Gamma \left(\frac{u}{\mu }+\frac{\lambda
   }{2}+1\right)^{\frac{2 u}{\mu }+\lambda } \Gamma \left(\frac{u}{\mu }\right)  \\ \nonumber
&&\Gamma \left(\frac{u+\mu
   }{\mu }\right) \Gamma \left(\frac{2 u-\lambda +\mu -2}{2 \mu }\right) \Gamma \left(\frac{2 u+\lambda
   +\mu +2}{2 \mu }\right) \times \\ \nonumber
&& \Gamma \left(\frac{u+\mu -1}{\mu }-\frac{\lambda }{2}\right)^{\frac{u-1}{\mu
   }-\frac{\lambda }{2}} \Gamma \left(\frac{\lambda }{2}+\frac{u+\mu -1}{\mu }\right)^{-\frac{2
   u+\lambda  \mu -2}{2 \mu }} \times \\
&& \Gamma \left(\frac{u+\mu +1}{\mu }-\frac{\lambda
   }{2}\right)^{\frac{u+1}{\mu }-\frac{\lambda }{2}} \Gamma \left(\frac{\lambda }{2}+\frac{u+\mu +1}{\mu
   }\right)^{-\frac{2 u+\lambda  \mu +2}{2 \mu }}\,,
\>
where $\zeta^{(a_1,a_2)}(s,z)$ denotes the $\frac{\partial^{a_1+a_2}}{(\partial s)^{a_1} (\partial z)^{a_2}}$ derivative of the Hurwitz zeta function.

The universal R-matrix construction also works for non-integer and even complex values of parameter $\lambda$, but it is significantly more complicated to find a compact crossing equation in this case.

\section{Conclusions and outlook}

The construction of R-matrices for integrable models is a long-standing problem. In this paper we have reformulated the universal R-matrix formula proposed in \cite{Khoroshkin:1994uk} for Yangians of simple Lie algebras and adapted it to the case of Lie superalgebras, and also to algebras for which the Cartan matrices contain non-integer elements. We have subsequently used this method to derive the R-matrices of the $\slnm$ and $\glnn$ algebras including their scalar factors. We found these factors to satisfy simple functional equations, which we conjecture to be the corresponding crossing equations. In the case $m=0$ we recover, up to a CDD-factor, the well-known $\mathfrak{su}(n)$ R-matrices with the correct dressing factors. For $\glnn$ algebras we find that for the non-canonical choice of the extension parameters $\mu$ and $\lambda$ the scalar factors exhibit intricate dependence on $\mu$ and $\lambda$. We evaluated the simplest such factor (for $n=1$) analytically and found the corresponding crossing equation. Based on this, we conjecture the period of the crossing equation for $\glnn$ to be equal $\mu \, n$. It would be useful as a cross-check to  derive these equations for arbitrary values of $\mu$ and $\lambda$ using the Hopf algebra implementation of the crossing symmetry. Indeed, we find that the Yangian automorphism generator of $\mathcal{Y}(\glnn)$ is shifted by $\mu n$ under the action of the antipode, whereas all other generators remain unaffected. We did not check explicitly that the application of the antipode to the R-matrix yields the crossing equation.

The universal R-matrix formula provides a compact plug-in type formula that should allow to evaluate the R-matrix including the corresponding dressing factor for \textit{any} algebra and \textit{any} representation. It should also work for any Dynkin diagram of a particular superalgebra. We checked this in Appendix \ref{sec:diff} for some Dynkin diagrams of $\slnm$ algebras. We thus believe that this approach is very powerful and should find several applications in the physics of integrable models. It would be interesting to study more systematically different representations of the Yangian of $\slnm$ and to evaluate the R-matrix on them. Furthermore, our formula for the universal R-matrix should also be valid for Yangians of other contragredient Lie superalgebras, such as the orthosymplectic Lie superalgebras $\alg{osp}(n|m)$. However, their representation theory seems to be much harder than for $\slnm$, so the evaluation of the R-matrix even on the fundamental representation of the simplest orthosymplectic algebra $\alg{osp}(1|2)$ seems complicated \cite{Arnaudon:2003ab}. As in our approach the elements of Cartan matrices may take arbitrary values, we believe that our abstract result for the universal R-matrix is also valid for the exceptional Lie superalgebra $D(2,1; \alpha)$. Drinfeld's second realisation of $D(2,1; \alpha)$ was defined for the related case of the quantum affine algebra in \cite{Heckenberger:2007ry},  whereas the Yangian of $D(2,1;\alpha)$, to our knowledge, has not been explicitly investigated in the literature. Such an explicit investigation of $Y(D(2,1;\alpha))$ would be very interesting from a mathematical point of view as well as for physical applications, e.g. in the AdS/CFT correspondence \cite{Matsumoto:2008ww}-\cite{Babichenko:2009dk}.

Finally, we believe that the abstract form of the R-matrix should also work in the case of centrally extended $\alg{sl}(1|1)$ and $\alg{sl}(2|2)$ algebras, which are of great relevance to the AdS/CFT correspondence \cite{Beisert:2005wm}-\cite{Beisert:2005tm}. However, even the classical case of the $\alg{sl}(2|2)$ loop algebra needed some important modifications \cite{Beisert:2007ty}, and the second realisation of the Yangian now involves some shifts by the central charge \cite{Spill:2008tp} rendering the evaluation considerably more complicated than in the rational $\slnm$ case. We would like to note that our method of constructing bases dual to the Cartan basis of the Yangian is from a technical point of view quite similar to methods of solving crossing equations for models with $\alg{su}(n)$ invariance \cite{Volin:2010cq}. This methods are known to work in the case of the centrally extended $\alg{sl}(2|2)$ algebra \cite{Volin:2009uv}. Thus, we believe that one can construct the Yangian double and the universal R-matrix also in this case. The universal R-matrix approach might help to answer the question whether physical magnons are built out of elementary excitations or not, see\cite{Rej:2007vm}.

\section*{Acknowledgements}

We would like to thank Nicolas Cramp\'e, Sergey Khoroshkhin, Alex Molev, Eric Ragoucy, Matthias Staudacher, Arkady Tseytlin and especially Dima Volin for interesting discussions and comments on the manuscript.
We also thank Peter Koroteev for inital collaboration. Adam Rej is supported by a STFC postdoctoral fellowship. Fabian Spill would like to thank the Deutsche Telekom Stiftung for a PhD fellowship.

\appendix

\section{The q-deformed Cartan matrix of $\slnm$} \label{sec:Aanditsinverse}

A natural step when considering the universal enveloping algebras is to introduce the $q$-numbers
\beq
n \to [n]_q = \frac{q^n-q^{-n}}{q-q^{-1}}\,.
\eeq
The $q$-bracket can be also expressed in terms of the Fibonacci polynomials
\beq
[n]_q = (-i)^{n-1} F_n (i\, [2]_q) = \prod^{|n|-1}_{j=1} \left( [2]_q - 2 \cos{\frac{ \pi j}{n}} \right)\,.
\eeq
It follows immediately from the definition that $[-n]_q=-[n]_q$ and $[n]_{1/q}=[n]_q$. The $q$-numbers obey the following relations
\beqa \label{fibrec}
&&[n]_q+[n+2]_q-[2]_q [n+1]_q=0\,, \\
&&[n]_q^2+[n+1]_q[n-1]_q=1\,,\\
&&[j n]_q = [n]_q \frac{\left[(j-1)n+2\right]_q-\left[(j-1)n-2\right]_q}{[2]_q}+\left[(j-2)n\right]_q\,,
\eeqa
for any $j \in \mathbb{Z}$.

The q-deformed Cartan matrix \eqref{def:cartnm} of $\slnm$ is given by

\[ \label{cartannm}
\cartnm(q)= \begin{pmatrix}
[2]_q &-1&0&\dots & &  & & & & \\
-1&[2]_q&-1&\dots &\vdots & & &0 & & \\
0&\dots & \ddots & -1&0& & & & & \\
\vdots&\dots&-1&[2]_q &-1 & & & & \\
0&\dots& 0&-1& 0& 1 & 0 &\dots  & & \\
& & & &1 &-[2]_q &1&0&\dots&\\
& & & &0 &1&-[2]_q&1&\dots&\\
& & 0& &\vdots & & & \ddots&1&\\
& & & &0 &\dots& &1&-[2]_q
\end{pmatrix}\,.
\]
As in the $q=1$ case one may reduce its determinant by using Laplace expansion twice to the determinants of the Cartan matrices of $A_n$. This results in 
\< \nonumber
\det A^{\alg{sl}(n|m)}(q) &=& (-1)^m(\det\cartq{n-1}\det\cartq{m} - \det\cartq{n}\det\cartq{m-1}) \\
&=& (-1)^m[n-m]_q\,.
\>
The determinant is thus proportional to the q-deformed dual Coxeter number. Furthermore, the matrix
\[
 C(q) := \left(\det \cartnm(q)\right) \left(\cartnm(q) \right)^{-1}
\]
consists only of integer powers of $q$ and $q^{-1}$. For $n \neq m$ the inverse of the Cartan matrix may be written in the following form

\beq \label{inversenm}
\left(\cartnm (q)\right)^{-1}= 
\begin{pmatrix}
a_{n+m-1,1}&\dots & \dots & \dots & \dots & \dots& \dots &\\
\vdots&\ddots&\dots &\dots & \dots& \dots & \dots &\\
a_{m+1,1}&\dots&a_{m+1,n-1}&\dots  & \dots& \dots& \dots &\\
 b_{m,1}
& \dots & \dots &b_{m,n} &\dots& \dots & \dots &\\ \vdots
& \ddots & \dots & \vdots  &c_{m-1,n+1}&\dots&\dots&\\ b_{2,1}
&b_{2,2} & \dots & \vdots  & \vdots & \ddots& \dots&\\b_{1,1}
& b_{1,2}& \dots & b_{1,n} &c_{1,n+1}&\dots &c_{1,n+m-1}&
\end{pmatrix} .
\eeq
Due to the fact that $\cartqb^{-1}$ is symmetric, we only present the lower half of the matrix. Clearly, three distinct blocks may be distinguished. The reader should note a different from usual numeration of the matrix elements, which allows more easily to understand the corresponding block structure. The relation between the both notations amounts to the shift $i \to n+m-i$. The elements of the three aforementioned blocks are given by
\beqa 
\label{abcnm1} a_{i,j}&=&-\frac{[2m-i]_q [j]_q}{[n-m]_q}, \quad m<i \leq n+m-1, \ 1 \leq j  \leq n+m-i \,,\\
\label{abcnm2} b_{i,j}&=&- \frac{[i]_q [j]_q}{[n-m]_q}, \qquad  1 \leq i \leq m,\ 1 \leq j \leq n \,, \\
\label{abcnm3} c_{i,j}&=&-\frac{[i]_q [2n-j]_q}{[n-m]_q}, \quad  1 \leq i \leq m-1,\ n+1 \leq j \leq n+m-i\,.
\eeqa
Please note that $[n-m]_q \, b_{1,1}$ is equal to $-1$, independently of $n$ and $m$. Moreover, the following relations between different matrix elements hold
\beq
a_{i,j}=b_{2m-i,j}\,, \qquad c_{i,j}=b_{i,2n-j}\,.
\eeq
The elements $a_{i,j}, b_{i,j}$ and $c_{i,j}$ obey the difference equations
\beqa
\label{dnm1}&&[2]_q \,a_{i,j} = a_{i+1,j}+a_{i-1,j}= a_{i,j+1}+a_{i,j-1}\,,\\
&&[2]_q \,b_{i,j} = b_{i+1,j}+b_{i-1,j}= b_{i,j+1}+b_{i,j-1}\,,\\
\label{dnm3}&&[2]_q \,c_{i,j} = c_{i+1,j}+c_{i-1,j}= c_{i,j+1}+c_{i,j-1}\,,
\eeqa
subject to appropriate boundary conditions
\beq
b_{i,j}=b_{j,i}\,, \quad b_{0,0}=0\,, \quad b_{1,0}=0\,, \quad b_{1,1}=-\frac{1}{[n-m]_q}\,.
\eeq
The relations \eqref{dnm1}-\eqref{dnm3} are essential for proving that \eqref{inversenm} is the inverse matrix of $A^{\alg{gl}(n|m)}(q) $.

In the case $n=m$ the q-deformation of the Cartan matrix \eqref{def:cartnn} is given by

\beqa
\nonumber && A^{\alg{gl}(n|n)}(q) = \begin{pmatrix}
[2]_q &-1&0&\dots & \dots &\dots  & \dots & \dots & \dots &0 \\
-1&[2]_q&-1&0 &\dots & \dots & \dots & \dots & \dots & \vdots\\
0&\ddots & \ddots & -1& 0& \dots & \dots & \dots & \dots &\vdots \\
\vdots&\dots&-1&[2]_q &-1 & \ddots & \dots &\dots &\dots & 0 \\
0&\dots& 0&-1&0&1&0&\dots  &\dots &\mu \\ \vdots
& \dots & \dots & \dots &1 &-[2]_q &1&0&\dots&0 \\ \vdots
& \dots & \dots & \dots &0 &1&-[2]_q&1&\ddots&\vdots\\ \vdots
& \dots & \dots & \dots &\vdots & \dots & \ddots& \ddots&1&\vdots\\ \vdots
& \dots & \dots & \dots &0 &\dots& \dots &1&-[2]_q&0
\\ 0
&\dots &\dots&0 &\mu&0&\dots &\dots&0&\lambda\mu
\end{pmatrix} 
.\\
\eeqa

The corresponding inverse matrix has a similar structure as \eqref{inversenm}, with the exception that the the last row and column are distinguished
\beq \label{inversenn}
\left(A^{\alg{gl}(n|n)}(q)\right)^{-1}= 
\begin{pmatrix}
a_{2n-1,1}&\dots & \dots & \dots & \dots & \dots& \dots &\vdots\\
\vdots&\ddots&\dots &\dots & \dots& \dots & \dots &\vdots \\
a_{n+1,1}&\dots&a_{n+1,n-1}&\dots  & \dots& \dots& \dots & \vdots \\
 b_{n,1}
& \dots & \dots &b_{n,n} &\dots& \dots & \dots & \vdots \\ \vdots
& \ddots & \dots & \vdots  &c_{n-1,n+1}&\dots&\dots& \vdots \\ b_{2,1}
&b_{2,2} & \dots & \vdots  & \vdots & \ddots& \dots&\vdots \\b_{1,1}
& b_{1,2}& \dots & b_{1,n} &c_{1,n+1}&\dots &c_{1,2n-1}& \vdots
\\d_{0,1}
& d_{0,2}& \dots & d_{0,n} &d_{1,n+1}&\dots &\dots&d_{0,2n}
\end{pmatrix} .
\eeq
The four parts of the matrix are defined as follows:
\beqa
a_{i,j}&=&\frac{[j]_q}{[n]^2_q} \,\left([i-n]_q [n]_q-\frac{[\lambda \mu]_q [2n-i]_q}{[\mu]^2} \right), \quad n<i \leq 2n-1, \ 1 \leq j  \leq 2n-i \,,\nonumber \\
b_{i,j}&=&- \frac{[i]_q [j]_q [\lambda \mu]_q}{[\mu]^2_q[n]^2_q} \,, \qquad  1 \leq i \leq n,\ 1 \leq j \leq n \,, \nonumber \\
c_{i,j}&=&-\frac{[i]_q}{[n]^2_q} \,\left([j-n]_q [n]_q+\frac{[\lambda \mu]_q [2n-j]_q}{[\mu]^2_q} \right), \quad  1 \leq i < n,\ n+1 \leq j \leq 2n-i\,. \nonumber \\
d_{0,j}&=& \left\{ \begin{array}{ll} 
\frac{[j]_q}{[\mu]_q [n]_q}\,, & \textrm{ $1 \leq j\leq n$}\\ 
& \\
\frac{[2n-j]_q}{[\mu]_q [n]_q}\,, & \textrm{ $n <j \leq 2n$}\end{array}\right.\,.
\eeqa

\section{Different Dynkin diagrams of $\slnm$} \label{sec:diff}

For simple Lie superalgebras there are several Dynkin diagrams with a different number of fermionic roots. This is in contrast to the case of simple Lie algebras, which have one unique Dynkin diagram and a corresponding unique Cartan matrix. As far as Yangian is concerned, it is clear that different Dynkin diagrams of the same algebra will lead to isomorphic Yangians. For the universal R-matrix this is difficult to prove due to the ordering issues. In the previous sections we have used the distinguished Dynkin diagram. Here, we extend these studies to the fermionic Dynkin diagram of $\slnm$ in the case of $m=n$ and $m=n\pm1$. 
\begin{figure}\centering
\setlength{\unitlength}{1pt}%
\small\thicklines%
\begin{picture}(120,40)(-10,-30)
\put(  5,00){\circle{15}}%
\put(  3.5,12){1}%
\put(  12,00){\line(1,0){21}}%
\put( 40,00){\ldots}%
\put( 57,00){\line(1,0){21}}%
\put( 85,00){\circle{15}}%
\put( 78,12){2n-1}%
\put( 0,-5){\line(1, 1){10}}%
\put( 0, 5){\line(1,-1){10}}%
\put( 80,-5){\line(1, 1){10}}%
\put( 80, 5){\line(1,-1){10}}%
\put(  5,-30){\circle{15}}%
\put(  3.5,-18){1}%
\put(  12,-30){\line(1,0){21}}%
\put( 40,-30){\ldots}%
\put( 57,-30){\line(1,0){21}}%
\put( 85,-30){\circle{15}}%
\put( 78,-18){2n-2}%
\put( 0,-35){\line(1, 1){10}}%
\put( 0, -25){\line(1,-1){10}}%
\put( 80,-35){\line(1, 1){10}}%
\put( 80, -25){\line(1,-1){10}}%
\end{picture}
\caption{The fermionic Dynkin diagrams of $\slnn$ and $\alg{sl}(n|n+1)$ .}\label{Dynkinferm1}
\end{figure}

To write down the fundamental representation in those cases, it is convenient to introduce a new basis $\tilde{V}_j$, $j = 1,\dots 2 n+1$. The case $m=n-1$ works similarly. We choose the basis vectors such that

\beqa\label{eq:fermbasis}
&&\tilde{V}_j \quad\textit{is odd for}\quad j=2 k, k\in \mathbb{N}\nonumber\\
&&\tilde{V}_j \quad\textit{is even for}\quad j=2 k + 1, k\in \mathbb{N}\,.
\eeqa
We will denote by $|i|$ the grading of the index $i$ such that it is even if $i$ is an even number, and fermionic otherwise. The $(n+m)^2$ matrices $E_{ij}$  are now even if $i+j$ is an even number, and odd otherwise. On the fundamental
representation of $\slnm$ the Chevalley-Serre basis is spanned by
\beqa
\csgen{H}_i &=& (-1)^{|i|}(E_{i,i}+E_{i+1,i+1}),\nonumber\\
\csgen{E}^+_i &=& E_{i,i+1},\nonumber\\
\csgen{E}^-_i &=& (-1)^{|i|}E_{i+1,i}\,,
\eeqa
and the additional Cartan generator for $\alg{gl}(n|n)$ is simply
\[
\csgen{H}_{2n} = E_{1,1}\,.
\]
Here, we do not introduce any rescaling or shift of the automophism as in the case of the distinguished diagram, as the effect would be exactly the same. 

We have checked that with this choice of the basis one gets on the fundamental representation exactly the same R-matrix up to the transformation of the basis vectors as for the distinguished diagram \eqref{eq:funR}. Henceforth, we believe that our results concerning the universal R-matrix are independent of the choice of the Dynkin diagram.

\end{document}